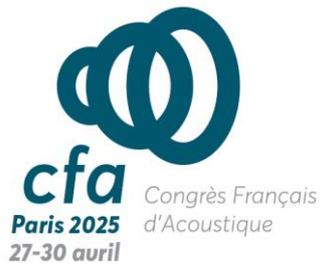

17ᵉ Congrès Français d'Acoustique
27-30 avril 2025, Paris

# Analyse de la décorrélation geste-son dans la pratique de la batterie électronique

M. Blandeau[a], E. Avril[a]

[a] Université Polytechnique Hauts-de-France, CNRS, UMR 8201 LAMIH, F-59313, Valenciennes, France

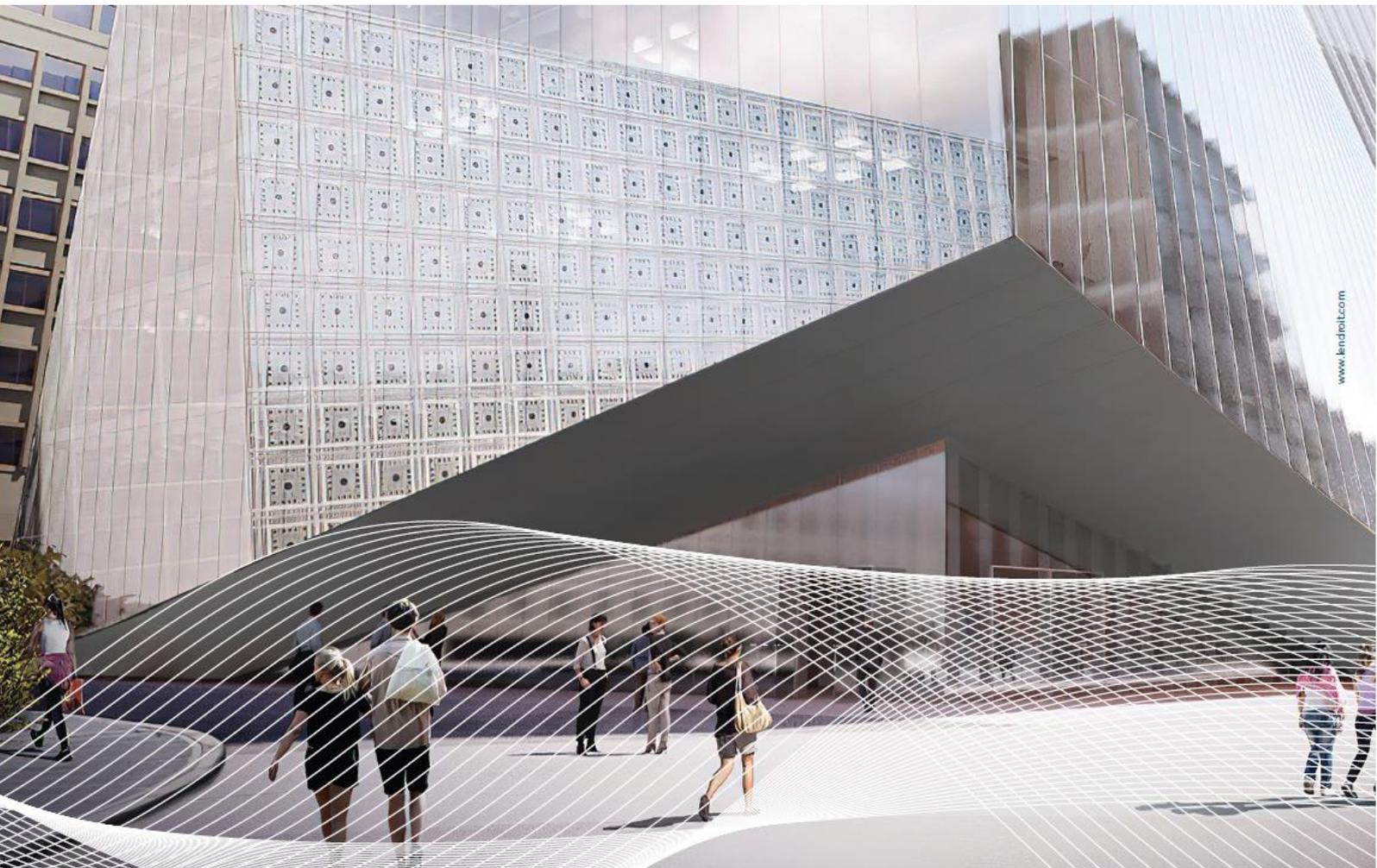

La batterie appartient à la famille d'instruments de musique dont la pratique, en amateur ou à haut niveau, est associée à un risque accru de troubles musculosquelettiques (TMS) particulièrement pour les membres supérieurs et le rachis lombaire. L'apprentissage de la batterie se fait en grande majorité sur des instruments acoustiques dont l'intensité sonore est proportionnelle à la force de frappe développée. Cette corrélation se retrouve perturbée lors de la pratique de la version électronique de l'instrument qui est souvent acheté par les musiciens cherchant à réduire le son produit (ex : jeu en appartement). L'objectif de cette étude est donc d'analyser si la pratique de batterie sur du matériel électronique entraînerait un changement dans la cinématique et le ressenti des batteurs. Dans ce but, plusieurs batteurs ont été recrutés pour réaliser des rythmes répétés à différentes nuances sur batterie acoustique et batterie électrique avec deux conditions sonores (son allumé et son éteint avec casque anti-bruit). Le son produit ainsi que la cinématique des membres supérieures ont été mesurés par capture du mouvement vidéo lors des frappes. Par ailleurs, des entretiens d'auto-confrontation ont été réalisés après chaque condition. Les batteurs, confrontés à des enregistrements vidéo de leurs actions, devaient décrire, expliquer et commenter pas à pas leur performance. Ces entretiens ont également servi à évaluer leur aptitude à maintenir une force de frappe constante. Un questionnaire a permis d'obtenir des informations subjectives sur leur ressenti. Les résultats ont montré une puissance sonore inférieure de la batterie électronique malgré une vitesse de frappe similaire, cette décorrélation geste-son pourrait expliquer l'augmentation des TMS chez les batteurs lors du changement d'un instrument acoustique à électronique.

## 1   Introduction

Les troubles musculosquelettiques (TMS) représentent les affections liées au travail les plus courants en Europe. Des douleurs lombaires aux douleurs musculaires, les TMS sont les principaux contributeurs aux services de réadaptation et aux années vécues avec un handicap (AVH), représentant environ 17 % de toutes les AVH dans le monde [1].

Le consensus scientifique actuel soutient que les sources des TMS sont multifactorielles, prenant en compte divers aspects physiques, psychologiques, sociaux et organisationnels des situations de travail [2]. Malgré ce modèle étiologique complet, les efforts de prévention se concentrent principalement sur la réduction des facteurs biomécaniques et physiologiques, car ce sont les plus faciles à mesurer.

Dans le domaine musical, les TMS passent souvent inaperçus, ce qui les rend plus difficiles à reconnaître et à accepter. Dans les orchestres, les blessures peuvent être perçues comme un signe de faiblesse, potentiellement nuisible à la carrière d'un musicien dont le revenu est directement lié à la capacité à se produire. Cette perception contribue à une « culture du silence », où les musiciens peuvent être réticents à parler de leur douleur ou à chercher de l'aide. Cette situation est aggravée par l'instabilité financière, où une blessure peut non seulement entraîner une perte de revenus, mais aussi compromettre les opportunités de carrières futures [3]. En conséquence, l'accès au soutien social et organisationnel, aux soins de santé appropriés et aux ajustements nécessaires des techniques de jeu est entravé. Les stratégies de prévention et de gestion des blessures dans ce domaine restent rares par rapport à d'autres professions physiquement exigeantes [4], [5].

La performance musicale est une activité physiquement exigeante, comparable à celle des personnes impliquées dans le travail manuel ou les activités athlétiques. Les musiciens, en particulier, les professionnels, sont exposés à divers facteurs pouvant entraîner des TMS liés à la pratique instrumentale. Tout d'abord, la nature physique du jeu musical, combinée aux mouvements répétitifs et aux longues périodes de jeu, expose ces musiciens à un risque particulièrement élevé de développer de tels troubles. Les TMS ont été étudiés au cours de la dernière décennie pour divers instruments comme les cordes grattées et frottées [6], [7], [8] ou le piano [9], [10], mais les résultats pour les batteurs n'étaient pas cohérents en raison de facteurs tels que la petite taille des échantillons, le mélange des groupes de population de percussionnistes dans les études et le manque d'inclusion de genres musicaux diversifiés. Les travaux pionniers d'Azar ont introduit le domaine du jeu de batterie et exploré l'importance des TMS des batteurs, les comparant notamment à des sports de haute intensité [11], [12], [13].

D'un point de vue ergonomique, la formation initiale, qu'elle se fasse seule ou accompagnée, à l'obtention du « geste métier » nécessaire à la bonne réalisation d'une activité dépend indéniablement de l'outil d'origine sur lequel ce geste est réalisé [14]. Il est logique de supposer que le changement d'outil, s'il n'est pas associé à une nouvelle formation puisse donner lieu à une inadéquation entre le geste métier et l'activité souhaitée et ainsi augmenter le risque d'apparition de TMS [15].

La pratique d'un instrument de musique doit également être analysée via le prisme du contrôle moteur et du paradigme de perception/action. Bien que le rythme soit communément associé aux expériences auditives découlant d'actions récurrentes produisant des motifs sonores cycliques, il est important de reconnaître que la perception du rythme s'étend au-delà de notre sens de l'ouïe. En effet, les motifs rythmiques peuvent être détectés et traités par divers autres canaux sensoriels. Ceux-ci incluent le sens tactile (système haptique), notre conscience de la position et du mouvement du corps (proprioception), le système visuel et le sens de l'équilibre et de l'orientation spatiale (système vestibulaire). Cette nature multisensorielle de la perception du rythme souligne les manières complexes et interconnectées dont nos corps interprètent et répondent aux stimuli rythmiques à travers différents domaines sensoriels [16].

Les différences de continuité temporelle entre les différentes modalités sensorielles peuvent inhiber l'intégration perceptive lorsque des stimuli appariés présentent des incohérences de continuité. Par exemple, Schutz et Lipscomb ont montré que la modification de la continuité d'un coup de marimba présenté visuellement — un mouvement d'impact court et saccadé par rapport à un mouvement d'impact plus long et fluide — entraînait des illusions de perceptions de la durée du son [17].

A la lecture de tous ses éléments l'objectif de notre étude est d'analyser l'évolution du mouvement de frappe de batterie lors de la décorrélation frappe/son qui peut se retrouver lors de la pratique de la batterie électronique afin d'étudier le potentiel risque du changement d'instrument sur l'apparition de TMS du membre supérieur chez le batteur.

## 2  Matériel & Méthode

### 2.1  Participants

Trois participants (2 hommes et 1 femme), tous majeurs, droitiers et ne présentant pas d'antécédents de blessure incompatibles avec la pratique de la batterie, ont accepté de prendre part à cette étude. Ils pratiquaient tous la batterie depuis au moins plus de 12 ans et avaient tous appris la pratique sur une batterie acoustique. A leur arrivée, le protocole d'acquisition leur a été présenté et après avoir donné leur consentement pour participer à l'étude, ils ont rempli un court questionnaire afin de connaître leur expérience et antécédant de pratique de la batterie acoustique et électronique.

### 2.2  Protocole

Chaque participant s'est vu donner une participation de caisse claire comprenant 5 mesures à jouer à un tempo de 100 battements par minute (cf. figure 1) La batterie acoustique se compose d'une caisse claire classique et la batterie électronique du pad de caisse claire d'une batterie numérique portable Yamaha Multi Pad DD-75.

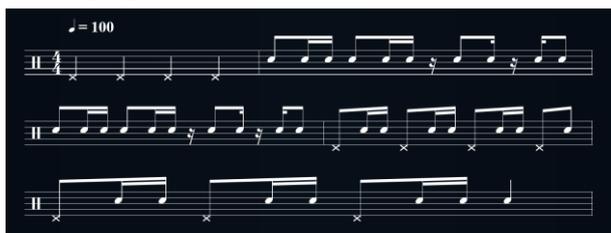

Figure 1 – Partition répétée dans les 8 conditions par chaque sujet.

Trois variables indépendantes ont été utilisées afin d'étudier l'impact du type de batterie, du retour auditif et de la nuance de jeu sur la pratique (cf. TABLEAU 1), l'ordre des conditions a été normalisé parmi les sujets afin d'éviter tout effet d'apprentissage.

TABLEAU 1 – Variables indépendantes

| Variable | Valeurs |
|---|---|
| Batterie | Acoustique / Electronique |
| Casque de protection | Avec / Sans |
| Nuance | Piano / Forte |

Lors de l'utilisation de la batterie en condition électronique, il était demandé aux sujets de régler le volume selon leur sensibilité afin de produire « un son similaire à ce que la batterie acoustique aurait produit pour une même frappe ».

Après chaque acquisition, une série de questions était posée afin de questionner l'expérience de jeu du sujet (cf. TABLEAU 2). Enfin un entretien d'auto-confrontation était réalisé en présentant aux sujets 2 vidéos successives en condition nuance forte et sans casque sur les batteries acoustique et électronique.

### 2.3  Acquisition et traitement de données

Deux outils de mesures ont été utilisés pendant les acquisitions, une caméra de smartphone de type iPhone 13 (iOS 18.3.1) enregistrant le sujet de trois-quarts face dans une vidéo cadencée à 60Hz et à une résolution de 1080p ainsi qu'un sonomètre PCE EM882 réglé en mesure de dBA maximal produit pendant l'acquisition.

TABLEAU 2 – Questions de performances.

| Question | Score |
|---|---|
| 1. Comment évaluez-vous votre force de frappe ? | Lickert 7 points |
| 2. A quel point vous sentiez-vous à l'aise en jouant ? | Lickert 5 points |
| 3. A quel point aviez-vous besoin de vous concentrer pour maintenir le rythme demandé ? | Lickert 5 points |
| 4. A quel point avez-vous besoin de vous concentrer pour maintenir la nuance demandée ? | Lickert 5 points |

Les données vidéos ont été traitées à l'aide du logiciel OpenPose [18] qui permet de détecter la pose 2D d'un ou plusieurs personnes dans une image (cf. Figure 4). OpenPose représente un des premiers logiciels dits de capture du mouvement « sans marqueurs » (*markerless*). Il est très répandu dans la communauté biomécanique (+32k étoiles sur le dépôt GitHub) et représente une base commune pour comparer les recherches cinématiques [19]. L'approche utilise une représentation non paramétrique, appelée Part Affinity Fields, pour apprendre à associer les parties du corps aux individus dans l'image.

Une fois le traitement des vidéos réalisés, les trajectoires des centres articulaires, exprimées en pixel, ont été extraites. Toutes absence de point de moins de 5 images ont été reconstruites en utilisant des splines cubiques. Enfin, les données de positions des points ont été filtrées à l'aide d'un filtre passe-bas de type Buttertworth d'ordre 4 et de fréquence de coupure de 6 Hz.

Trois segments anatomiques ont été identifiés,
- Le bras droit (vecteur allant de l'épaule au coude droit)
- L'avant-bras droit (vecteur allant du coude au poignet droit)
- La main droite (vecteur allant du poignet à la 1ᵉʳᵉ phalange de l'index droit)

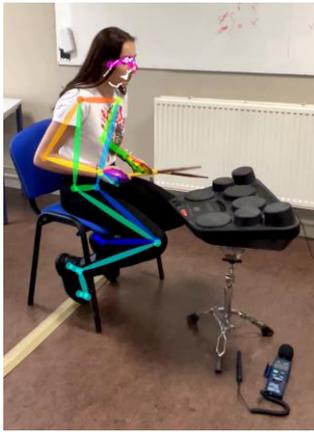

Figure 3 – Suivi de centres articulaires par le logiciel OpenPose.

L'angle du coude (resp. poignet) a été calculé dans chaque acquisition comme la distance angulaire relative entre les segment bras (resp. avant-bras) et avant-bras (resp. main). Enfin, la vitesse angulaire a été extraite pour chaque angle.

Une interface a été programmée (cf. Figure 3) afin de détecter les instants des trois frappes principales du bras droit (la 1ᵉʳᵉ frappe des mesures 2 et 3 ainsi que la dernière frappe de la 5ᵉ mesure). Cette interface combinait l'affichage du signal vidéo et d'image et permettait l'identification à l'aide d'un curseur temporel représentant la position temporelle dans l'acquisition.

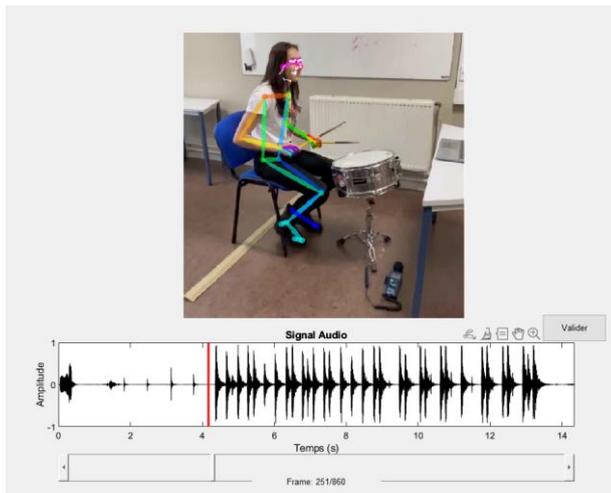

Figure 2 – Interface pour l'identification des frappes.

Les valeurs de vitesses angulaires de coude et poignet ont été sélectionnées comme les maxima locaux sur ces signaux à plus ou moins 1/5ᵉ de seconde autour des instants identifiés.

La comparaison statistique des données quantitatives issues des 3 conditions d'acquisition a été réalisée par la méthode du T-test. Les données qualitatives ont quant à elles été analysées par un test de rang signé de Wilcoxon. Toutes les opérations d'extraction et de traitement des données ainsi que les analyses statistiques ont été effectuées sous Matlab (version R2022b, The MathWorks, Natick, MA, USA).

# 3 Résultats

## 3.1 Mesures sonores

La Figure 2, présente les différentes valeurs des niveaux sonores maximaux mesurés. Des différences significatives ont été retrouvées en fonction du type de nuance ($p<0.05$) ainsi que du type de batterie ($p<0.001$). La présence ou l'absence de retour auditif n'a pas produit de différence sur le niveau sonore.

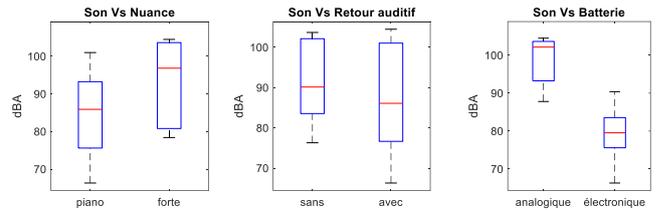

Figure 4 – Evolution du niveau sonore par rapport aux conditions expérimentales.

## 3.2 Mesures cinématiques

On observe sur la Figure 5, l'effet de la nuance, du retour auditif ainsi que du type de batterie sur les mesures cinématiques des sujets. Une différence significative a été calculée concernant la vitesse angulaire du poignet ($p<0.001$) et du coude ($p<0.001$) avec des vitesses mesurées jusqu'à 3 fois plus rapide entre la nuance piano et forte. Les paramètres de retour auditif et type de batterie n'entraînent pour leur part pas de différence significative.

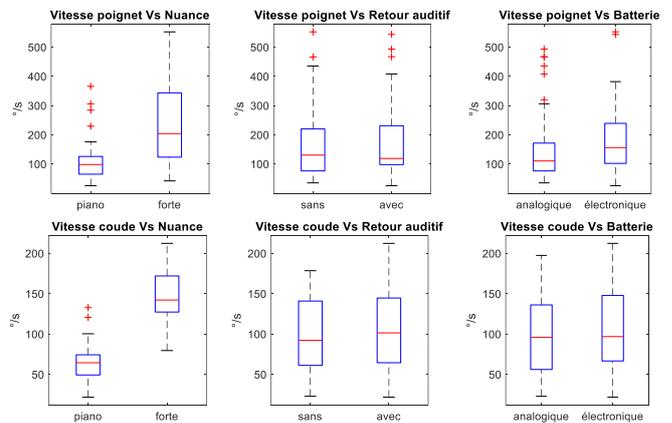

Figure 5 – Evolution des paramètres cinématiques des articulations du poignet (ligne du haut) et du coude (ligne du bas) par rapport aux conditions expérimentales.

## 3.3 Questionnaires et entretiens

On observe sur la Figure 6, l'effet de la nuance, du retour auditif et du type de batterie sur les scores des échelles de Lickert aux quatre questions posées entre chaque condition de jeu. On observe un effet significatif de la nuance sur les scores à la première question (force de frappe, $p<0.001$) et deuxième question (sentiment d'être à l'aise, $p<0.05$). Le retour auditif et le type de batterie n'ont pas apporté de changements significatifs aux scores des quatre questions. Les entretiens d'auto-confrontation ont permis d'identifier chez le sujet le plus expérimenté une différence de ressentie de jeu entre les batteries acoustiques et électroniques. Les gestes réalisés étaient perçus comme de trop grande amplitude avec un besoin de précision plus important et une fatigue ressentie plus grande en fin de concert/répétition lors de jeu sur batterie électronique. Les sujets avec moins d'expérience de jeu sur batterie électronique n'ont pas remonté de quelconque différence de jeu entre les deux types de batterie exception faite du « retour de frappe » (mouvement de baguette après le choc) différent.

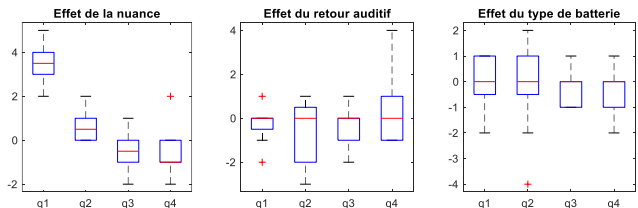

Figure 6 – Différence de score sur les échelles de Lickert pour les quatre questions posées en fonction des conditions de nuance, de retour auditif et de type de batterie.

## 4 Discussion

L'objectif de cette étude est d'analyser l'impact du retour auditif sur la frappe de batterie définie en termes de cinématique articulaire, de puissance sonore et de ressenti. Tout d'abord ce travail présente certaines limitations expérimentales, les principales étant le nombre trop faible de sujets lié à une difficulté de recrutement et le fait que la batterie soit limitée à la seule caisse claire même si ce type d'étude a déjà été réalisé par le passé [20]. L'utilisation du sonomètre en mode maximal présente également le défaut de ne pas savoir quelle frappe a été spécifiquement associée au son mesuré. De plus, étant donné que les sujets frappaient dans leurs baguettes pendant l'acquisition, il est possible que certaines acquisitions en nuance piano aient été surestimées. L'alternative consistant à analyser le signal audio des vidéos n'a pas été retenue car le signal sonore saturait très souvent en nuance forte. Enfin concernant l'étude cinématique, le logiciel OpenPose n'est plus mis à jour et peut souffrir d'erreur d'estimations non-corrigées, de futures études consisterons à utiliser des logiciels plus récents comme Sports2D [21] ou Pose2Sim [22].

Les résultats concernant la puissance sonore peuvent paraître évidents dans le lien entre le son et la nuance mais on remarque néanmoins un effet du type de batterie alors qu'il avait été spécifiquement précisé aux sujets de régler le volume sonore de la batterie de sorte à produire le même son qu'avec la batterie acoustique. Un des sujets a signifié sa frustration du rendu sonore final après avoir été obligé de tourner le potentiomètre de réglage sonore au maximum. Ce décalage entre le son produit par la batterie électronique et acoustique pourrait, en l'absence d'un système de sonorisation supplémentaire, pousser des musicien à frapper plus fort sur le pad et ainsi accroître la sollicitation mécanique de leurs articulations, augmentant ainsi le risque de blessure [3].

Concernant la partie cinématique, l'utilisation de l'analyse du mouvement markerless est à notre connaissance une première dans le domaine de la batterie. Il est difficile de capturer le mouvement de batteurs (en plus largement de musiciens) en conditions dites écologiques (i.e. habituelles, naturelles), notamment parce que les approches traditionnelles basées sur la capture du mouvement par marqueurs entravent les mouvements naturels et sont très sensibles aux conditions environnementales [22]. De manière similaire à l'analyse de la puissance sonore, le jeu en nuance forte a augmenté la vitesse articulaires du membre supérieur ce qui implique une augmentation de la sollicitation mécanique et donc le risque de TMS [12].

Enfin, la combinaison de l'analyse cinématique avec une approche ergonomique de l'activité a permis d'approfondir considérablement nos résultats. Cette approche combinant analyse quantifiée du mouvement, questionnaires et entretien a précédemment démontré par le passé un intérêt certain dans l'analyse du risque de TMS dans l'activité professionnelle [23] ou même dans la proprioception pendant la rééducation du membre supérieur [24]. La performance musicale, transportant le musicien dans un état de transe [16] ou de cognition musicale incarnée [26], est associée positivement à des mouvements amples, souples et dépourvus de contraintes physiques associées à de la frustration de la part du musicien. Aussi, la nécessité ressentie de précision augmentée lors de la pratique de la batterie électronique nécessite de solliciter la totalité des capacités motrices du membre supérieur. Ce « dégel » des degrés de libertés se heurte à ce que Bernstein avait identifié [25] comme une stratégie d'apprentissage permettant de simplifier la commande motrice et ainsi gagner en efficience tout en réduisant le risque de blessure.

## Conclusions

Les risques de TMS dans la pratique musicale doivent encore être étudiés afin d'améliorer la santé et la formation des musiciens de demain.

Les résultats de cette étude soulignent l'importance d'une approche ergonomique et biomécanique pour prévenir les TMS chez les musiciens. De futurs études se focaliseront sur l'analyse du cycle complet de la frappe ainsi que dans la perception globale de la performance de jeu et la différence mécanique liée à la nature du matériau de façon située dans l'activité

## Remerciements